\documentclass[12pt]{article}
\usepackage{amsfonts,amssymb,amsmath,amscd}

\newcommand{\Tr}{{\rm Tr}}

\newcommand{\hc}{{\hat c}}

\newcommand{\tsigma}{{\tilde\sigma}}

\newcommand{\cV}{{\cal V}}

\newcommand{\be}{\begin{equation}}
\newcommand{\ee}{\end{equation}}

\title{A remark on worldsheet fermions and double-scaled matrix models}
\author{Anton Kapustin\\
{\it California Institute of Technology, Pasadena, CA 91125, U.S.A.}}

\begin{document}

\begin{titlepage}

\maketitle

\begin{abstract}
We provide a heuristic explanation for the emergence of worldsheet fermions in the continuum limit
of some matrix models. We also argue that turning on Ramond-Ramond flux confines the fermionic degrees of freedom of the Ramond-Neveu-Schwarz formalism.
\end{abstract}

\end{titlepage}

It has been realized recently that certain noncritical Type 0 superstring backgrounds admit
a nonperturbative description by means of double-scaled matrix models~\cite{TT,hat,hat2}. 
Namely, Type 0B backgrounds
with $\hc\leq 1$ are described by two-cut Hermitian matrix models, while Type 0A backgrounds with 
$\hc\leq 1$ are described by complex (and in general rectangular) matrix models. 
The evidence given in Refs.~\cite{TT,hat,hat2}
is circumstantial: it has been shown that many observables in the matrix and worldsheet descriptions agree.
It would be very interesting to find a direct argument for the emergence of worldsheet fermions
and worldsheet supersymmetry in the continuum limit of these matrix models. On one hand, it is far 
from clear that this goal is feasible: after all, Type 0 strings are described by GSO-projected 
superconformal field theories, which do not have any fermionic states in the spectrum. In other words, 
one may regard worldsheet fermions and worldsheet supersymmetry as an artefact of a particular continuum 
description of Type 0 strings, and it is entirely possible that the matrix models discussed in 
Refs.~\cite{TT,hat,hat2} 
discretize some other continuum description of the same string theories. On the other
hand, the RNS formalism is the only known continuum description of noncritical superstrings, and 
it is natural to try to interpret the matrix model fatgraphs in RNS terms. 
The main benefit of such an interpretation
would be a qualitative understanding of the effects of the Ramond-Ramond (RR) 
flux on the worldsheet degrees of freedom
in the RNS formalism. We will argue below
that turning on a RR flux leads to a confinement of the fermionic degrees of freedom. 

In this short note we present a heuristic picture of how the fermions emerge from the matrix models.
Previous discussions of this issue can be found in Refs.~\cite{KM,three}.
We will show that our picture explains some features of the matrix-model/superstring duality in
a natural way. At the very least, this should be regarded as evidence that the matrix model fatgraphs
provide a discretization of the RNS worldsheet.

We will mostly focus on a special case: Type 0B superstring with $\hc=0$. The corresponding matrix model
is a large-N Hermitian matrix integral with a symmetric double-well potential. The saddle point with
equal numbers of eigenvalues in the two minima of the potential corresponds, in the simplest double-scaled
limit, to the simplest Type 0B backgrounds: $N=1$ super-Liouville theory coupled to $bc$ and $\beta\gamma$
ghosts. The Liouville field and the $bc$ ghosts describe the metric on the Riemann surface;
in the discretized approach we expect them to be replaced by random fatgraphs. On the other hand,
local supersymmetry on a lattice is problematic, so we will treat the superpartner of the Liouville
field, the Mayorana fermion $\psi$, as well as the $\beta\gamma$ ghosts, as matter fields. On our very
heuristic level of discussion, we will ignore the $\beta\gamma$ ghosts and try to ``see'' the 
Mayorana fermion $\psi$ in the structure of the fatgraphs of the Type 0B matrix model.

When constructing the diagrammatic expansion of the matrix model partition function, it is important
to expand about the correct vacuum. In the case of the double-well potential $V(x)$ with minima at $x=a$
and $x=-a$, this means that the
eigenvalues must be localized near $\pm a$, and (for vanishing RR flux) the
number of eigenvalues in the two minima must be the same. To accomplish this it is natural
to partially gauge-fix the $U(N)$ symmetry by requiring the matrix to be block-diagonal, with
blocks of size $M\times M$, $M=N/2$~\cite{Ketal}. As usual, this leads to ghosts, which in the
present context are $M\times M$ matrices $C_1,C_2,B_1,B_2$ with Grassmann-odd entries. The gauge-fixed
action is~\cite{Ketal,three}
\begin{multline}\label{action}
S(X,Y,B,C)=\Tr\left(V(X-a)+V(Y-a)-2a(B_1 C_1-B_2 C_2)\right.\\
\left. +B_1 X C_1 - B_2 X C_2 + Y B_1 C_1 - Y B_2 C_2\right).
\end{multline}
Here $X$ and $Y$ are $M\times M$ matrices describing eigenvalues localized near $a$ and $-a$, respectively.
The residual gauge symmetry is $U(M)\times U(M)$. The diagrammatic expansion of the corresponding
matrix integral leads to fatgraphs with the following additional structure. There are several types
of vertices: the ones coming from $V(X)$, which we color black, the ones coming from $V(Y)$, which
we color white, and four kinds of vertices involving the $BC$ matrix ghosts. $X$ and $Y$ propagators
cannot connect vertices of different color, but they can terminate at ``ghost''
vertices. Also, the ghost number is conserved, so
the ghost propagators (which can be represented by double-lines, just as the $X$ and $Y$ propagators)
form closed loops. These loops separate regions with only black ($X$) vertices from regions with only
white ($Y$) vertices. Thus each Feynman diagram is a random fatgraph discretizing a Riemann surface plus
some loops on it which separate black and white domains. This picture was recently discussed in 
Ref.~\cite{three}, where a similar description of the Type 0A complex matrix model was also given.

Recall now that the Mayorana fermion can be discretized by means of the Ising model. More precisely,
the continuum limit of the Ising model with periodic boundary conditions at the critical temperature is described by a massless Mayorana fermion with a Type 0B GSO projection. The necessity of the Type 0B GSO projection can be seen intuitively as follows~\cite{Polyakov}. Each state in the Ising model can be represented by a collection of domains on the lattice where the spins are pointing ``up''. In the complement
of these domains the spins are pointing ``down''. The fermion loops are essentially the boundaries
of these domains. The change from the spin variables to the fermionic variables is possible
because knowing the boundaries of the domains completely specifies the spin configuration. One
potential problem with this argument is that on a Riemann surface of nonvanishing genus 
there exist closed loops which
are homologically nontrivial and therefore are not boundaries of domains. Contributions of such
loop configurations must vanish or be removed ``by hand''. The Type 0B projection achives precisely this.
For example, in the case of a torus, imagine a loop which winds once around the $a$-cycle, 
which we will regard as the ``time'' direction. By cutting the torus
along the $b$-cycle, we see that this contribution to the partition function comes from a state with 
a single fermion. GSO projection removes this configuration from consideration.

In view of the above, it is tempting to conjecture that the worldsheet fermion $\psi$ is the fermionization
of the black and white domains on the fatgraphs of the Type 0B matrix model. That is, the $BC$ ghost loops
can be identified, in the continuum limit, with $\psi$ loops. 

We stress that at present this is merely a suggestive analogy, because the Ising model coupled to gravity is certainly not equivalent to the noncritical 0B superstring. In the remainder of this note we will give a couple of examples showing that our heuristic picture has some explanatory power, so perhaps it is more
than a mere analogy. A proper derivation of the RNS worldsheet theory from the matrix model fatgraphs remains
to be found; it would have to explain where the $\beta\gamma$ ghosts come from. Hopefully, this can be
achieved by studying in detail the difference between the fatgraphs of the 0B matrix model and the
Ising model coupled to gravity.

It was argued in Ref.~\cite{hat2} that the response of the Type 0B string to RR flux is qualitatively
different depending on the sign of the super-Liouville coupling constant $\mu$. 
This coupling multiplies the term
\begin{equation}\label{sL}
\psi_+\psi_- e^{\phi}
\end{equation}
in the continuum action. For $\mu>0$ every term in the perturbative expansion of the free energy is
analytic in the RR flux $q$, while for $\mu <0$ half-integral powers of $q$ appear. Thus only for
$\mu > 0$ is expansion in the number of RR insertions well-defined. In the case
of 0B string with $\hc=0$ the RR vertex operator in question has the form
$$
\cV=e^{-\frac{\varphi+\bar\varphi}{2}} \sigma,
$$
where $\varphi$ and $\bar\varphi$ are the right and left-moving bosonized superconformal ghosts,
and $\sigma$ is one of the two twist operators for $\psi$ (the other one being removed by the GSO
projection).

A similar distinction arises in the Ising model picture. Recalling that $e^{\phi}$ is the conformal scale of the worldsheet metric, we see that the term Eq.~(\ref{sL}) is the mass term for the fermion $\psi$. Thus the super-Liouville coupling $\mu$
corresponds to the deviation of the Ising temperature from criticality. Ignoring the $\beta\gamma$
ghosts, $\cV\sim\sigma$ is simply a twist operator for the fermion $\psi$. There are actually two fermionic
twist operators, $\sigma$ and $\tsigma$.
In the Ising language, $\sigma$ is the spin operator which lives on the vertices of the lattice, while 
$\tsigma$ is the so-called disorder operator which lives on the vertices of the dual lattice.
The disorder operator is defined as follows~\cite{KC}. One draws a contour $L$ on the dual lattice which begins at the point where we wish to insert a disorder operator and runs off to infinity 
(or to the insertion point of another disorder operator). 
Then one flips the sign of the Ising coupling constant on each bond of the original
lattice intersected by $L$. It is easy to see that the choice of $L$ is immaterial. It is also easy
to see that in the ordered (low-temperature) phase the disorder operator creates a defect line. Thus
the correlation function of two disorder operators in the low-temperature phase will be of order 
$e^{-\ell/a}$ where $\ell$ is the distance between the insertion points and $a$ is the lattice cut-off. 
In other
words, in the low-temperature phase the disorder operator is completely screened at macroscopic scales.
On the other hand, in the high-temperature phase it is the spin operator which has exponentially decreasing correlators. In fact,
the Kramers-Wannier duality exchanges the high and low-temperature phases and the spin and disorder
operators. If we consider Ising spins on a compact surface, Kramers-Wannier duality is broken.
For example, with periodic boundary conditions 
the state corresponding to the disorder operator is not in the spectrum of the critical Ising model, 
for any $\mu$. This truncation is equivalent to the 0B GSO projection on the Mayorana fermion.
As explained above, the deviation of the Ising temperature from criticality is proportional to $\mu$.
Thus the spin operator $\sigma$ is screened for one sign of $\mu$ and has a long-range order for the other
sign.

To determine which phase of the Ising model corresponds to which sign of $\mu$, we recall~\cite{hat2}
that $\mu >0$ corresponds to the situation when the resolvent of the matrix model has two cuts, while
for $\mu <0$ they merge into a single cut. Since the cuts of the resolvent indicate the location
of the eigenvalues of the matrix, we conclude that for $\mu>0$ the eigenvalues are well localized
near the minima of the potential, and therefore the fatgraphs are in the ordered (low-temperature)
phase. 

To provide more evidence for this identification, consider turning on the RR flux for $\mu >0$.
On one hand, the RR vertex operator is basically the twist operator
for the fermion, which we have identified with the spin operator of the Ising model. Adding a spin operator
to the Ising action amounts to turning on a magnetic field. It is well-known that this perturbation
is relevant and gives in the contunuum a rather complicated (but for $\mu=0$ still integrable~\cite{Zamo}) 
massive field theory. The Mayorana fermion is not among the elementary excitations of this theory,
because for it is {\it confined} by the magnetic field. Indeed, the magnetic field favors spins of a particular orientation, which means that in the ground state all spins have the same orientation. The contribution of a ``droplet'' of spins with the opposite orientation to the partition function is suppressed by $e^{-hA}$, where $A$ is the area of the droplet and $h$ is the magnetic field. This means that the contribution of a fermion loop will be exponentially suppressed by the area of the domain that it bounds, 
i.e. fermions are confined. 

On the other hand, in the matrix model turning on the RR flux
is achieved by making the numbers of eigenvalues localized in the two minima unequal~\cite{hat,hat2}.
In order to have $M_1$ eigenvalues in one minimum of the potential and $M_2$ eigenvalues in the other minimum, one gauge-fixes the matrix so that it is block-diagonal
with blocks of size $M_1\times M_1$ and $M_2\times M_2$~\cite{Ketal,three}. 
The gauged-fixed action is given by the
same equation Eq.~(\ref{action}), where now $X$ and $Y$ have sizes $M_1\times M_1$ and $M_2\times M_2$,
respectively, the $B$-ghosts are rectangular matrices of size $M_2\times M_1$, and the $C$-ghosts are rectangular matrices of size $M_1\times M_2$. The RR flux is proportional to
$M_1-M_2$. Suppose $M_1>M_2$. Then, according to the usual t'Hooft counting rules, each face of a black domain will contribute $e^{M_1}$ to the partition function, while each face of a white
domain will contribute $e^{M_2}$. Thus the contribution of white domains will be suppresed by a relative
factor of order $e^{-(M_1-M_2)A}$, where $A$ is the area of the white domains. Therefore for $M_1\neq M_2$
the $BC$ ghosts are confined, in agreement with our identification of $\psi$ loops with the $BC$ loops.
This also confirms that the two-cut ($\mu>0$) phase of the matrix model corresponds to the
ordered (low-temperature) phase of the Ising spins.

We end this note with the following remark. In the limit of large RR flux it appears that Type 0B $\hc=0$
string theory reduces to a noncritical bosonic string. Indeed, one of the cuts in the eigenvalue density recedes to infinity in this limit, so that only one cut remains near the maximum of the potential,
and this situation should be describable in the continuum by a bosonic string. Our heuristic picture
provides the following explanation for this. As one turns on the RR flux, while keeping the string 
coupling fixed, the fermions become confined by the worldsheet magnetic field and decouple. We speculate
that more generally turning on RR flux confines fermionic degrees of freedom in the RNS formalism,
leading to a bosonic string in the limit of large RR flux.

The author would like to thank Jongwon Park for useful discussions and Juan Maldacena for comments. 
This work was supported in part by the DOE grant DE-FG03-92-ER40701.

\end{document}